# DA VINCI – ARCHITECTURE-DRIVEN BUSINESS SOLUTIONS

AN ORIGIN WHITE PAPER

ORIGIN B.V.

AUTHOR(S) : H.A. (Erik) Proper

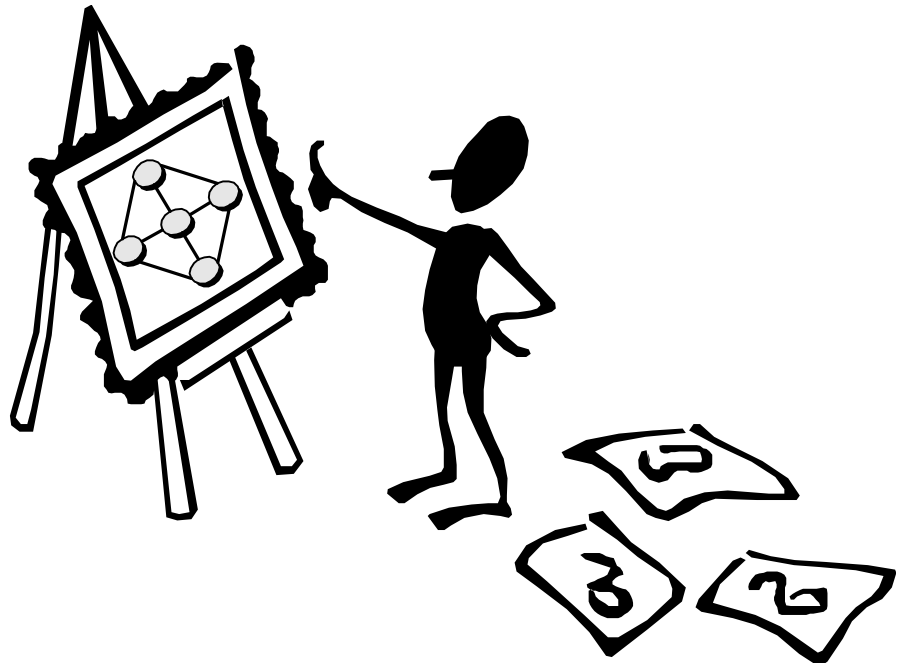

*There is a theory which states that if ever anyone discovers exactly what the Universe is for and why it is here, it will instantly disappear and be replaced by something even more bizarre and inexplicable. There is another theory which states that this has already happened.*

*THE RESTAURANT AT THE END OF THE UNIVERSE DOUGLAS ADAMS, PAN BOOKS LTD.*

# DA VINCI – ARCHITECTURE-DRIVEN BUSINESS SOLUTIONS

| | | |
|---|---|---|
| DOCUMENT NUMBER | : | |
| VERSION | : | 1.0.0 |
| SOURCE | : | Origin/Architecture Modelling |
| STATUS | : | White Paper |
| DOCUMENT DATE | : | 26 October 1999 |
| NUMBER OF PAGES | : | 38 |

| | | | |
|---|---|---|---|
| RELEASED BY | : | | INITIALS: |

Contents

# DA VINCI – ARCHITECTURE-DRIVEN BUSINESS SOLUTIONS

version: 1.0.0

## List of changes

| VERSION | DATE | DESCRIPTION | AUTHOR |
|---------|------|-------------|--------|
| 0.1.0 | 01-08-1997 | First version: 'The Hei Release' | H.A. Proper |
| 0.1.6 | 19-08-1997 | 'Hei sessie' input from Hans Bossenbroek and Bard Papegaaij | H.A. Proper |
| 0.1.15 | 29-09-1997 | Changed the focus from Picasso only to Architecture Modelling in general. Use of DA VINCI as the name. | H.A. Proper |
| 0.1.21 | 10-10-1997 | Restructuring of contents c.f. DA VINCI framework | H.A. Proper |
| 0.1.30 | 12-11-1997 | Several improvements after further discussions with Tim Barrett, Araminte Bleeker, Guus Boot, Hans Bossenbroek, Bert Kessels, Paul Veerkamp, and Martin Zomer | H.A. Proper |
| 0.2.0 | 16-11-1997 | Second version: 'The Hei-2 Release' | H.A. Proper |
| 0.2.16 | 07-02-1998 | Stronger focus on the *what* of architecture-driven application development | H.A. Proper |
| 0.2.32 | 18-06-1998 | Due to comments by Ronald Hetharia, Niels Klinkenberg, and Bard Papegaaij, the focus of DAVINCI is now shifted from 'application development' to 'business solutions positions DAVINCI as a way of thinking for the development of business solutions under architecture. | H.A. Proper |
| 1.0.0 | 23-06-1998 | We have reached a point where the ideas behind DAVINCI need to be tested out in practice. Before continuing on the 'theoretical' road, some practical experience is needed. | H.A. Proper |

## List of figures

# DA VINCI – ARCHITECTURE-DRIVEN BUSINESS SOLUTIONS

version: 1.0.0

1  Introduction

This document has emerged out of Origin's past experiences with architecture-driven application development ($AD^2$), and the need to further formalise and consolidate these experiences.

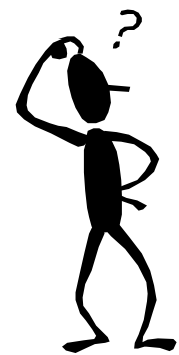

1.1  Architecture-driven application development

The $AD^2$ related developments range in scope from the actual design and implementation of applications, to the development of a long-term vision of an organisation's business activities and IT support required. The main concern of $AD^2$ is the development of applications to support an organisation's business activities, by considering the entire context of the applications.

Within Origin's Business Applications & Solutions (BAS) service line, $AD^2$ came first into being with the introduction of Picasso [1, 2]. The focus of Picasso is *application design under architecture*, where the phrase *under architecture* refers to the fact that applications are to be designed in harmony with some *context architecture* (e.g. the business processes).

While the focus of Picasso is clearly on design of applications only, it has slowly evolved to include more and more business aspects as well. The alignment of business and IT has always been core to the original Picasso way of thinking. Practice has shown that achieving this alignment requires a proper understanding of the business and associated processes. This has lead to extensions of Picasso that include more business issues. In [3] an example of a business oriented extension to Picasso can be found.

On the realisation side, Tim Barrett [4] has proposed the Technical Reference Architecture (TRA) as an architecture-driven approach to the realisation of IT solutions. A more recent iteration of this work can be found in [5], where an approach to component based transactional solutions is proposed. In Mondriaan [6], a Small-Talk based implementation framework has been proposed with similar goals..

1.2  Mission

Picasso, Mondriaan, and TRA are examples of methods which each cover a part of $AD^2$. Now, at the eve of 1998, it is time to consolidate and reflect on the $AD^2$ related developments so far. The different methods for $AD^2$ are shaping up to provide an integrated methodological framework for the development of architecture-driven business solutions (ABS).

The aim of this document is to identify Origin's way of thinking regarding ABS. As such, this document limits itself solely to the *what* of ABS. The aim is not to define detailed work procedures (the *how*). To honour a past artist, architect, and inventor, the name 'DA VINCI' has been selected as a name for this way of thinking. DA VINCI is firmly based on Origin's past experiences, and related approaches as documented in literature. The longer-term goal of DA VINCI is to arrive at a more mature approach to ABS.

Origin is not the only organisation working on approaches to ABS or $AD^2$. Other players are developing architecture-driven approaches as well. In defining Origin's way of thinking regarding ABS, this document also aims to differentiate DA VINCI from the other players. In a separate section, we will briefly review some of the alternative approaches.

1.3  Audience

This document is intended for anyone who is involved with ABS related activities. However, due to its high level of abstraction it may not be the right document for a first introduction to this field of work. This document aims at completeness with regards to the way of thinking. It is not tuned for one particular audience.

1.4  Structure of the document

Origin's motivations for architecture-driven application development are given in section 2 in terms of some *key drivers* and relevant *value statements*. To better position the different aspects involved in methods for ABS, section 3 briefly discusses the anatomy of a method. This anatomy will then be used as the guiding structure for the body of this document.

Sections 4 through to 6 each discuss different aspects of the methodological framework needed for architecture-driven business solutions. Section 4 focuses on the *way of thinking*, while the other sections discuss some of the consequences for different (two) aspects of the methodological framework. As the aim of this document is not to define the actual methods, the focus is on *what* issues only. Section 7 concludes the document. Finally, a dictionary of terminology used in this document can be found in section 8.

## 2 Motivation

This section is concerned with the motivation for DAVINCI. We start by discussing Origin's key drivers for DAVINCI. We believe that these drivers are also its important selling points:

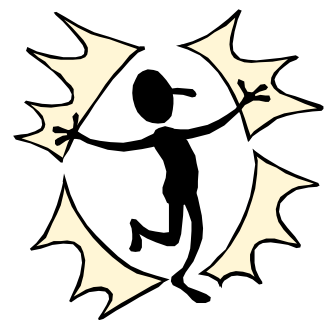

- Evolution is a constant
- Driven by needs; enabled by technology
- Stakeholders aware
- Results oriented; not role oriented
- Controlled evolution
- Knowledge creation
- Elegance above all

In the remainder of this section, each of these drivers will be discussed in more detail.

### 2.1 Evolution is a constant

The conditions under which most organisations currently operate exhibit a high degree of dynamism. A dynamism that forces them to become more flexible: "evolve or die". Reduced protectionism, deregulation of international trade, de-monopolisation of markets, privatisation of state owned companies, increased global competition, cross-border merges, the emergence of new trade blocks, the introduction of common currencies, are all aspects of this increasingly dynamic business environment. In other words, the domains in which organisations operate continuously evolve, and the speed at which they evolve is still increasing; evolution is a constant.

An organisation can deal with this evolution in a variety of ways. While some may try and continue their business 'as usual', others may choose to embrace the new developments and try to exploit their potential to its fullest. Neither approach is a guaranteed way to success or failure. Embracing new developments too early may lead to organisational chaos and decline, while waiting too long may result in missed business opportunities.

In [7, 8] a very elaborate discussion can be found of the changes in context and culture that are occurring inside organisations as well as in their environments. Tapscott proposes an architectural approach as a solution to make the needed changes to the organisational structure and in particular IT.

For Origin, it is particularly interesting to look at the role of IT in this arena. The Gartner Group has published numerous reports on architectural issues. In [9], and more recently in [10, 11], the Gartner Group identified the need for a shift in IT paradigms. This shift is needed to meet the need for more flexibility of the IT architecture. A shift is needed from a *technology-centred* to a more *application-centred* approach. Evolving business needs translates directly to evolving requirements on underlying applications. Taking a more application-centred approach will lead to IT architectures that are more attuned with the evolving business needs.

While IT should indeed support organisational processes and changes, at the same time it should not rule them. The role of IT can be characterised as a position between two undesirable extremes. On the one extreme, IT can completely restrain organisational change. In the other extreme, it can (try to) bring about organisational changes in a pace which is far too high. The latter situation can be compared to putting a Ferrari engine in a VW Beetle, while the driver is used to easing along at maximum speeds of 80 to 100 km/h.

In the first extreme, IT appears to smother any organisational change. This seems to be one of the dilemmas of IT. While it is reasonable to say that advanced IT should lead to revolutionary improvements in the flexibility and effectiveness of organisations, it is actually the same IT that slows organisational change by anchoring the organisation to rigid systems. Usually, these systems are built as monoliths that are the embodiment of the, at that moment, prevailing culture and structure of the organisation. In the past this has already led to so-called 'legacy systems'. Continuing to build systems in the same way will only create the legacy systems of tomorrow.

In the second extreme, IT will drive an organisation to a pace of change that goes beyond the speed its organisational fabrics can manage. Some organisations are just not ready to cope with the profound changes brought about by a technology push. For example, organisations that have only just reached a stage at which they are confident with the use of some basic IT to automate the processing of orders might simply not survive a quick move towards electronic commerce.

The general message is that IT should *empower* a business with the means to go out and seek new challenges. These new demands on IT in the new and rapidly evolving world, can be summed up by quoting [12]:

> "In the past an architecture was really the design of a system that had been created to meet specific application needs. In the new business environment, organisations have little idea what their application needs will be in two, let alone five or ten years. Consequently, we need architectures that can enable the exploitation of unforeseen opportunities and meet unpredictable needs."

One word of caution seems to be in order though. Flexibility and architecture should not be seen as synonyms. Firstly, situations exist where concessions to flexibility in favour of efficiency, say, are quite acceptable. Secondly, the systems we now call 'legacy systems' have an underlying architecture as well. It is just that their architecture may not always meet our current ideas about proper architecture.

Evolution in an architecture-driven context can be compared to an evolutionary software development process, where the development never ends. Evolutionary software development itself originates from prototyping based approaches. The idea of the prototyping approaches was to incrementally develop a prototype of the system/applications. The aim of this prototyping process was *not* to incrementally build the final system, but rather to incrementally elicit the complete set of *requirements* for the system [13]. Prototyping is therefore mainly employed as a learning process. After the prototyping phase, the actual system can be built using the completed set of requirements. The prototyping approach resulted from the observation that in a significant percentage of cases, as argued in e.g. [14], system requirements cannot be established correctly and completely in advance.

When dealing with large systems, the time between finishing the final prototype and finishing and delivery of the final system will be considerable. In those cases, it is quite likely that the delivered system does indeed meet the 'old' requirements but that, as the organisation has evolved further, the system does not meet the current requirements [15]. Software development in such cases is like shooting at a moving target.

This has sparked the idea to stretch the prototyping principle beyond the development phase. By setting up the prototype in such a way that it can be used as the final product, new requirements can be met by making further increments/changes to the 'prototype'. This has led to the evolutionary approaches for software development, an approach which has been introduced under different names. In [16] it is referred to as *evolutionary delivery*, in [17] as *the spiral model*, and in [18] as *iterative enhancement*. In evolutionary approaches there is no essential difference between the development and maintenance phases. In both phases, the system evolves by a sequence of adjustments, which have become necessary due to changes in the requirements. An architecture-driven approach would enhance this even further by taking the evolution of the business-context into consideration as well, in addition to the evolution of the underlying IT applications.

## 2.2 Driven by needs; enabled by technology

Ideally, business strategists should be able to focus solely on the development of a business, while IT plays the role of a catalyst. IT should therefore be the enabler of change, which gently furthers changes; neither pushing nor restraining an organisation's evolution. In [19], the Gartner Group identifies an increasing trend to move away from technology-driven architectures to more business-driven architectures. In that same publication, the Gartner Group also makes the following observation:

> "Although many organisations are beginning to put application architectures in place these architectures remain largely focused on, and constrained by, the lower levels of technology infrastructure."

Tapscott [7] argues that organisations move between three levels of maturity. According to Tapscott, IT should be one of the essential enablers of this process. In Figure 1, taken from [7], these levels are depicted. By redesigning their business processes, organisations will be able move to a situation in which teams can perform better. This development leads to *high-performance business teams*, where the focus is on the use of IT to enable teams to perform business functions. This requires a shift away from a hierarchical view on organisations to a more team based view. The next shift involves the integration of the business teams to an integrated whole, leading to an *integrated organisation*. The role of IT in these cases, is increasingly one of being an enabler; *from cost centre to profit centre*. By linking their systems to other organisations, in particular customers and suppliers, an organisation can finalise the paradigm shift, and become an *extended enterprise*.

In [20], the Gartner Group supports this view by stating:

> *"By 2001, several iterations of technical architectures will allow the evolution of enterprises to a constituency-driven inter-enterprise business environment (0.7 probability)."*

All these shifts and developments pose new requirements on the IT function. These shifts, however, will not come from IT alone. As Michael Hammer argued in his seminal paper: *Re-engineering work: don't automate, obliterate* [21], improving the flexibility and efficiency of business processes is more than just using IT to make it go faster. IT is only part of the answer, it should not be looked upon as the sole bringer of solutions, but rather as an enabler. Investments in a re-engineering of the business and better business-IT alignment will be needed just as well. However, organisational change should be business-driven, and not (solely) IT-driven.

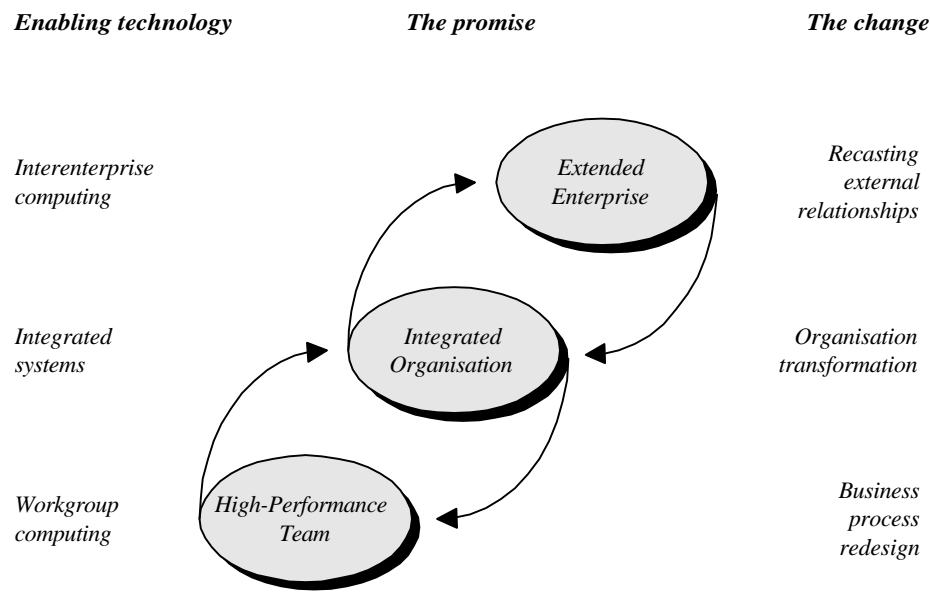

*Figure 1: The enabling effect of information technology*

It should be noted that a statement such as *"driven by needs; not by technology"* goes much further than organisational or business issues alone. For example, the needs of humans in terms of their use of technology plays a crucial role as well. A well-balanced borderline between human work and automated work carries the promise of a true synergy. The outcry for IT that *enables* organisational change rather than *inhibits* it is clearly louder than ever. IT, and associated application development, should therefore be driven by the needs of the business.

## 2.3   Stakeholders aware

In [22] it is argued how a detailed stake holder analysis at the outset of any *domain engineering* project is pivotal to the success of the project. The arguments used, however, directly apply to an organisation's architecture-driven development efforts. The discussion below is therefore strongly based on the discussion provided in [22].

Organisations adopting an architecture-driven approach to the development of business solutions may be faced with several challenges in establishing a strategic context for the project. There can be many motivating factors behind an organization's decision to take an architecture-driven approach. The business context and rationale for doing may not always be explicitly documented at the start of any architectural effort. There may be "hidden agendas" for various stakeholders, and conflicts that are not amenable to open discussion.

A sustainable approach to architecture-driven development requires multiple stakeholders whose interests form a healthy symbiosis. This configuration can not be manipulated by management mandates; the stakeholders have to 'come to the table' willingly. Clearly, this is not just a matter of documenting a set of 'given' project objectives. Setting publicly agreed upon objectives at the beginning of the project based on the interests of project stakeholders can help the project to focus on realistic objectives that will motivate stakeholders to buy into the project.

There may be both enablers and barriers for the various stakeholders. For example, some stakeholders may feel that their jobs are threatened by the new developments. For instance, someone with a 'guru' status with regards to some pre-existing legacy application may feel threatened by ideas about a new architecture replacing 'his' legacy system. Composing a complete picture of stakeholders and objectives means painting in the 'lights and the shadows', not just the conditions that appear to support the project.

## 2.4 Results oriented; not role oriented

In traditional software development methodologies, such as SDM [23], there is a strong focus on the distinct phases in a development project. This has lead to prototypical roles such as *information analyst*, *functional designer*, *technical designer*, *software engineer*, etc. Each of these roles comes with its own magical mist of 'experience'. The danger of these developments is that it may result in *islands of accountability* and experience driven *heroism*. People with a specific role in a certain phase will only feel (and be kept) accountable for 'their' role in 'their' phase, while their 'experience' with that particular phase and role may blind them in understanding the adjacent phases.

Additionally, it can be observed that the communication of results between analysts, designers, and software engineers is not always as good as it should be. Models passed on from business analysts to application designers often leave more questions unanswered than answered in the eyes of the designers. At the same time, the designs passed on by the application designers to the software engineers are quite often only used as 'inspiration' rather than as a proper design. In these cases, the transitions between each of the phases act as moats between the islands of accountability. Whenever such miscommunications occur, the final applications are most likely not like the business managers expected them to be, leading to a bickering over who is to blame.

## 2.5 Controlled evolution

Decisions concerning the future IT or business architecture may have a profound impact of the future of an organisation as a whole. Therefore, it is of the utmost importance for the organisation's managers to have control over these decision-making processes. Trade-offs should be made explicit by making proper evaluation, and new system investments should be prioritised.

What seems to be required are 'handles of control' that empower business people with a means to indeed take control over these architectural decisions. However, business people should not have to understand all technical intricacies of architectural issues, but should be able to base their decisions based on their business impacts. This would require some form of a common language that is understood by both business and IT people, and can be used to give direction to architectural developments.

Tapscott [7] proposes *architecture principles* as a way to provide these handles of control. In his terms, architecture principles are statements of *preferred architectural direction or practice*. Keen [8] refers to these principles as *policies*.

## 2.6 Knowledge creation

Developing advanced business solutions requires a constant tuning of the required skills, techniques, and methods. We need to operate in a way in which we constantly learn from our mistakes and successes. If a project fails, we need to know why, so we can avoid repeating the same mistake in the future. If a project is successful, we also need to know why to be able to make similar projects in the future succeed as well.

In the field of software development, these needs have already lead to developments such as:

- generalised forms of (knowledge) re-use, for instance, the knowledge creating software enterprise [22],
- continuous improvement efforts, such as the self-optimising level of the Capability and Maturity Model (CMM) [24].

More general theories about knowledge creating (or learning) organisations can be found in e.g. [25-27].

Putting in place a process of constant learning is of strategic importance to Origin. Anyone, including Origin's competitors, can 'learn' a new method. Once methods for architecture-driven development have been developed and well documented, anyone (depending on the person's intellectual capacities) can learn how to do it. However, the strategic power for Origin will lie in its ability to constantly tune and refine these methods based on day-to-day experience with development under architecture.

## 2.7 Elegance above all

We should never lose sight of the fact that any good design should simply be 'elegant'. It should simply look good! Be it the design of a building or the design of an information system, designing architectures is first and foremost a creative process.

Guidelines and principles can indeed be used to *falsify* candidate architectures, but they can most often not be used to automatically generate one completely. There is always a subjective aspect of elegance involved. Some architectures are 'right', simply because they look and feel good!

Finally, consider the following loose translation of a statement by Gustave Eiffel:

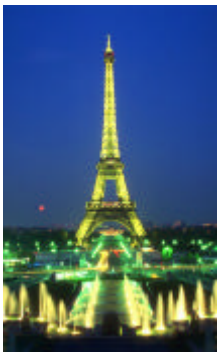

*"Should it be assumed that, since we are engineers, we are not interested in beauty, and that for the sake of the durability and strength of our constructions we do not strive for elegance?*

*Isn't it the case that the strength of a construction nearly always presupposes an almost hidden form of harmony?*

*The first principle of architectural aesthetics dictates that the essential forms of a monument are determined by a perfect alignment to its aims"*

    *Gustave Eiffel, 1887*

## 2.8 Discussion

The interesting thing about the above drivers is that at times they may seem almost trivial and sound as 'common sense'. One may indeed wonder what is so new about them. We certainly do not claim them to be new! While we may say that these drivers are not new, we can also observe that application development in practice does not deal with the issues raised. Then why don't we start working this way? For some reason, most of the issues raised and their possible solutions remain 'theory'.

In the past and in the present, many potential improvements to the way we develop applications/business-solutions have simply been dismissed as 'theory'. These potential improvements were consequently sacrificed to the gods of pragmatism. In old and proven books on software engineering [13, 28, 29] one may already find valuable lessons on the construction of software, that would have helped us in developing more maintainable applications. Then why is it that these construction principles have not been applied in practice? Is it because adding clear abstraction layers to software is theory? Is it because using the principles of abstract data-typing to hide implementation details of dates and currencies from other parts of the programs is theory?

One may come up with different reasons why potential improvements of the development process are simply dismissed as theory, tt may have to do with the level of professionalism of people involved, it may be because the improvements do not apply to the practical situations at hand, there may not be enough time to change the development process or to acquire new skills required., etc.

The ideas presented in this report essentially form a theory for architecture-driven business solutions. This report does not aim to discuss how this theory can be applied and *tested* in practice. This is left to the developers of the different service products that are to be developed in this arena. However, we should realise that the biggest challenge we are faced with will most likely not be the definition of these service products, or even selling the services based on these products to our clients. What will be the bigger challenge is to convince the people who will do the actual development that this is not just theory. The issues involved may be summed up by the following question:

> *Is Origin ready mature enough for architecture-driven business solutions?*

3   The anatomy of a method

As stated in section 1, DA VINCI is a *way of thinking* with regards to the development of architecture-driven business solutions. It does not attempt to define a specific method, but merely discuss what methods for ABS should entail. To make the relationship between DA VINCI and these actual methods clearer, we will start by discussing a general anatomy of a method.

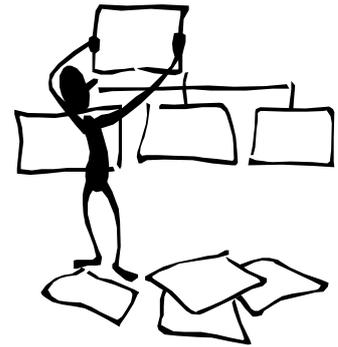

Figure 2 depicts a framework that provides a structured view of development methods. This framework is based on the framework originally presented in [30, 31]. Further inspiration from the field of Knowledge Creating Software Enterprises [22], in particular the plan-enact-learn idiom, has resulted in some refinements. In the resulting framework, a method is dissected into the following aspects:

- *Way of thinking.* This verbalises the assumptions and viewpoints of the method on the kinds of problem domains, solutions and modellers. This notion is also referred to as 'Die Weltanschauung' [32], or simply 'philosophy' [33].

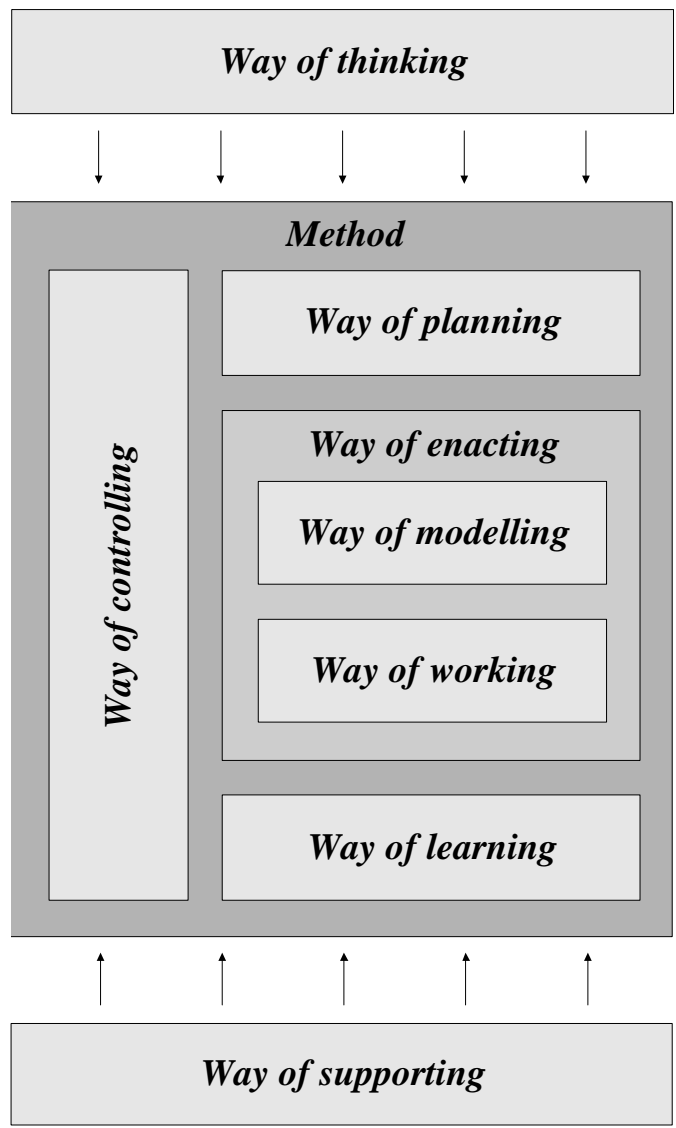

*Figure 2: Aspects of a method*

- *Way of controlling.* These are the typical project management aspects of system development. It includes such aspects as human resource management, quality and progress control, and evaluation of plans, i.e. overall project management [34, 35].

- *Way of planning.* This involves the (detailed) planning of the actual work that needs to be done. This includes such things as a project plan, but it may (possibly even should) go further than that. It may also contain plans for explicit learning, knowledge creation, and measuring.

- *Way of enacting.* This is where 'things' get done, i.e. the production oriented part. It can be further subdivided into:

    - *Way of modelling.* A description of the underlying modelling concepts together with their interrelationships and properties. In plain English, the actual techniques used to express the models, precise definition of deliverables, etc.

    - *Way of working.* This aspect focuses on the structure of the actual development process. It defines the possible tasks, including sub-tasks, and ordering of tasks, to be performed as part of the development process. It furthermore provides guidelines and suggestions (heuristics) on how these tasks should be performed.

- *Way of learning.* Here we are concerned with the way learning and knowledge creation is facilitated. It should provide answers to questions such as: *"How can we learn from past experiences? How can the method be refined to reflect new experiences?"*.

- *Way of supporting.* The way of supporting of a method refers to the support that may be provided by (possibly automated) tools. These tools can be CASE-Tools, project management tools, knowledge management tools, etc.

In practice, usually methods only cover part of these aspects. For example, SDM [23] mainly focuses on a way of controlling, while Origin's Rainbow™ procedure provides an umbrella for the way of controlling. The original version of ER [36] merely provides a way of modelling. A method like ORM/NIAM [37], on the other hand, provides in addition to a way of modelling a very detailed, almost dogmatic, way of working. Note that what we usually refer to as a 'technique' only provides a way of modelling, possibly supported by a way of supporting.

DA VINCI defines Origin's *way of thinking* with regards to ABS, and discusses the consequences of this way of thinking for the other aspects of methods for ABS.

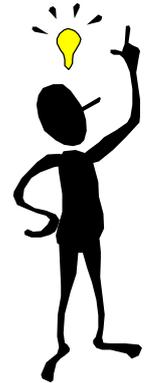

## 4 Way of thinking

This section forms the core of this report in the sense that it defines the DAVINCI way of thinking, revolving around the solution-context alignment and development under architecture. The remainder of this report is 'just' a consequence.

### 4.1 Strategic alignment

The relationship between business and IT is a major issue for architecture-driven business solutions. At least two of the drivers discussed in section 2 are involved:

1. Evolution is a constant
2. Driver by needs; not by technology

The importance of a good alignment between business and IT is discussed by both Tapscott & Caston [7] as well as Keen [8], and they are certainly not the first in doing this. In [38], Parker and Benson already discussed the issue of 'strategic alignment' between business and IT strategies. They argued that information technology planning and strategic considerations are part of the circular process as depicted in Figure 3. A distinction is made between the business domain on the one side, and the technology domain on the other side. Business planning drives how an enterprise will be organised, which should on its turn drive the technology planning to support the business. Technology planning leads to the discovery of further opportunities for future uses of technology, which will influence further business planning and strategy.

Parker and Benson also recognise the fact that this cyclic process may not work on an enterprise wide scale. Enterprises usually don't operate in a way that supports a 'monolithic' view of their information systems. However, they also argue that these cycles can be specialised to a specific line-of-business, or a specific business unit. In other words, the cycle of Figure 3 can be applied to smaller, more focussed, scopes within an enterprise.

The views of Parker and Benson were refined further by Henderson and Venkatraman [39]. On the importance of alignment between business and IT strategies, they argue:

> *We argue that the inability to realise value from IT investments is, in part, due to the lack of alignment between business and IT strategies of organisations.*

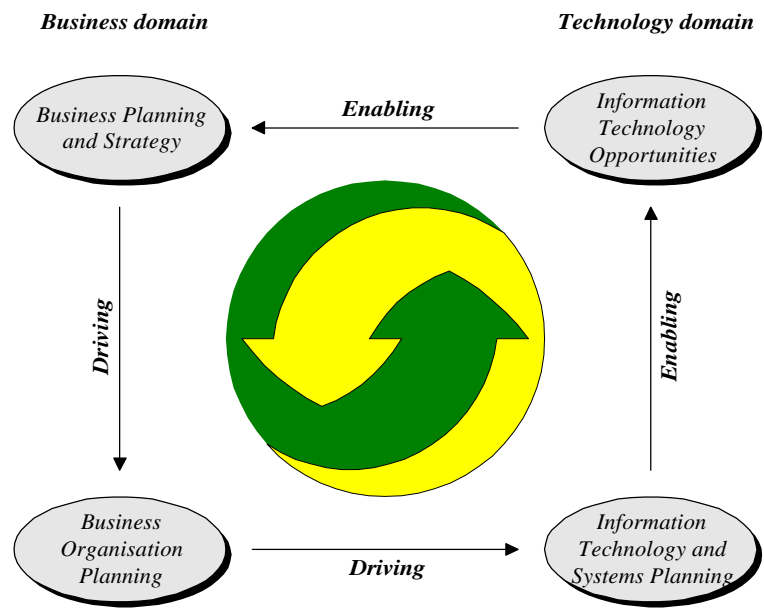

*Figure 3: Strategic alignment*

They also conclude:

*Strategic alignment is not an event, but a process of continuous adaptation and change.*

Below we will see how alignment between business and IT is not something that is limited to the strategic level only. Alignment between business and IT is needed on all levels, including tactical and operational levels.

Having a better understanding of the relationship between business strategies and IT strategies and their realisation would also allow for the development of reference models and reference solutions fitted to the needs of specific strategic archetypes. To further illustrate this point consider Figure 4, which is taken from [40]. Treacy and Wiersema argue that organisations should try and focus on one of the three following extremes:

1. Product leadership – These organisation aims to provide the best and/or most innovative products. An example would be Nike.

2. Operational excellence – These are typically organisations, which strive to provide a basic level of service in the most efficient way. McDonalds would be a prototypical example of operational excellence.

3. Customer intimacy – Organisations, which are customer focussed and aim to provide (complete) solutions for these customers. An example of such an organisation would be Rolls Royce (the car manufacturer).

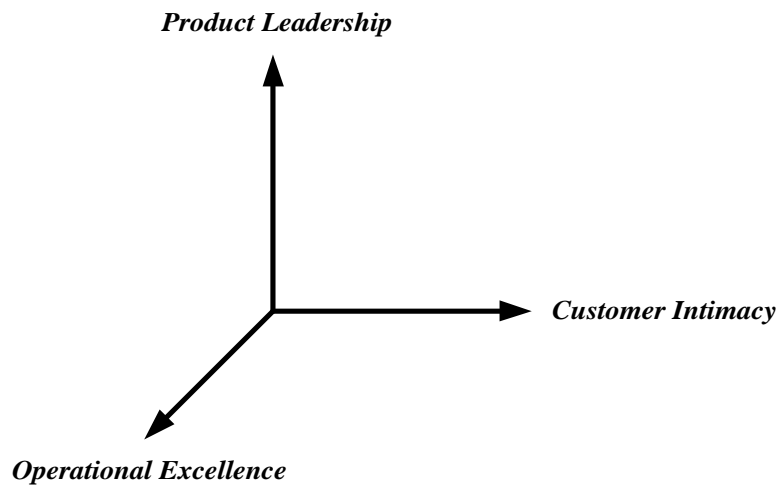

*Figure 4: Strategic focus*

When developing a business strategy, organisations will most likely focus on one of these three extremes. The IT strategy will be a resultant of the business strategy and somehow will reflect this latter choice. Even more, depending on the choice, certain IT components will come into focus or move out of focus. For example, organisations aiming for customer intimacy will probably be looking at data-warehousing, call-centres, and WWW technologies. An organisation striving for operational excellence will most likely cling to proven technologies, and shy away from modern middle-ware technologies. Understanding such stereotypical effects of business strategies on IT strategies and realisation would allow Origin to develop patterns of 'technology constellations' for these different stereotypes.

4.2   What is architecture?

Most readers will be familiar with the term architecture as such, in particular in the context of buildings. In the IT field, derived terms such as system-architecture, application-architecture, business-architecture, etc., are also widely used. It is not the aim of this section to delve into the philosophical backgrounds of architecture and try to establish what exactly it is. We only intend to clarify its meanings in the context of DA VINCI.

Suppose we find ourselves talking about the architecture of some building we are looking at. During this conversation, we can use the term architecture in two distinct, yet related, ways:

1. We can refer to the design of this particular building. For instance, how beautiful or ugly it is.
2. We can refer to the style of architecture used to design this building. For example, Medieval architecture, or Roman architecture.

According to the dictionary, [41], *architecture* indeed has two important interpretations:

1. Architecture is the art of planning, constructing, and designing buildings
2. The architecture of a building is the style in which it is designed and constructed.

We could therefore say:

> architecture = discipline + style

The cyclic process depicted in Figure 3 can be seen as the 'architecture process'. This is where an architectural discipline would come into play. The architectural style will refer to the construction of the actual models and other components used in the realisation of business solutions.

## 4.3 Development under architecture

Now that we have a better understanding of architecture, we can define what we mean by 'development under Figure 5, we have depicted a generic view on development under architecture. The actual development consists of three major phases: conception, definition, and realisation. The conception phase is aimed at understanding the business issue. This is where we may also find a stake holder analysis, and the traditional requirements analysis. During the definition phase, a solution to the business issue is defined, which is then realised during the realisation phase.

It can also be seen how the development process receives input from two directions: the context architecture, and the architectural experience. The experience may consist of tacit experience, which is in the minds of the people involved in the process, and formalised experience. Examples of formalised experience are:

- *Guidelines.* There may be written (and unwritten) guidelines of how to arrive at good architectural designs.
- *Reference architectures & patterns.* It is well known that for specific classes of problems generic solutions can be constructed that can be tailored to situations that are more specific. This may range from the re-use of generic patterns and components, to the application of complete reference architectures.

In addition to the architectural experience, the development process will receive input from the context architecture. The solution needs to be devised such that it fits within a certain architectural context; i.e. *under architecture*. This context may be the current context of the applications under development, but can equally be a long-term vision. Compare this to the way the architectural design of a new office building fits within the zoning plan of the area in which the building will be erected. At the same time, the interior decorator of the office building needs to ensure that the interior is in balance with the building itself. A context architecture may therefore be as simple as a set statements expressing a preferred architectural direction, or as detailed as a set of models.

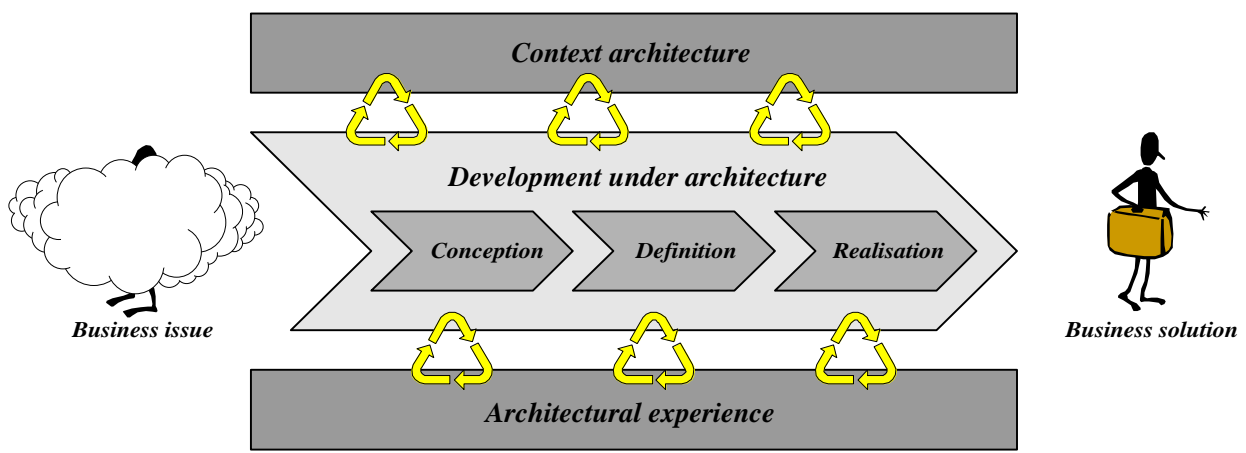

*Figure 5: Development under architecture*

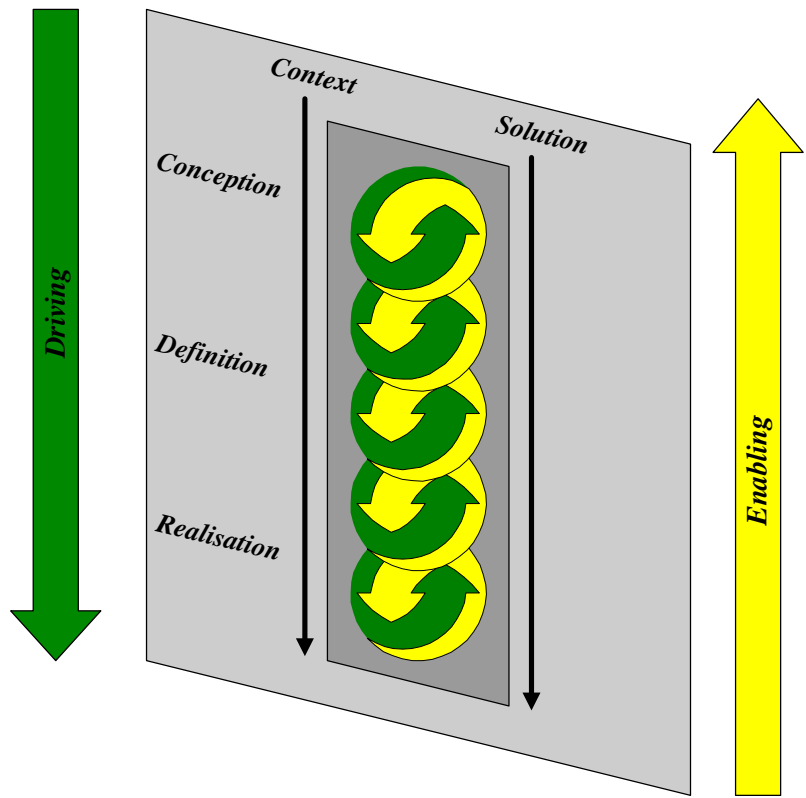

*Figure 6: Solution-context alignment*

In Figure 5, it is also illustrated that the development process not only receives input while it progresses, it also yields output. When developing new solutions, we will gain more experience, come up with new guidelines, and obtain more patterns and (partial) reference solutions. By learning from its experiences in the development of solution, the development organisation itself may become a learning organisation [22, 25-27].

The context architecture itself may also change as a result of the ongoing development process. While developing a solution, the enabling effects of parts of the solution may trigger the context in making further changes. It may eventually also lead to a change of the original problem definition as determined in the conception phase. This cyclic process could be referred to as solution-context alignment. In Figure 6, we have depicted the resulting cyclic process of driving and enabling between solution and context. When developing IT solutions in a business context, this alignment process corresponds to the earlier discussed business-IT alignment. As an example, consider a group of companies that introduce a customer loyalty scheme:

*Example 1:*

>   *Consider a group of companies that try to improve the loyalty of their customers. In other words, they want to ensure that once a customer has started purchasing goods (be it petrol, airline tickets, or groceries) they will continue to do so next time, and not wander of to competitors.*
>
>   *This group of companies decides to collectively introduce an 'air-miles' scheme by which customers can collect bonus points towards free tickets by purchasing goods from one of the participating companies. Such programs exist at least in the Netherlands (AirMiles), and Australia (FlyBuys), but are bound to exist elsewhere as well.*
>
>   *Now consider the development of such an air-miles scheme, and associated IT infrastructure. Suppose one would opt for the use of smartcards to administer points collected and redeemed. When using a smartcard, this opens up the opportunity to use the card for additional purposes. It may actually trigger someone in saying "Hey guys, what if we use these air-miles smartcards to introduce a one-card-does-it-all payment system".*

> *One could quite conceivably envisage a system whereby the smartcard knows (encrypted of course) about a customers accounts, including bank accounts, credit cards and Petrol cards, has a debit functionality, and stores a the payment preferences of the customer.*
>
> *This means that customers may be able to use the card when buying groceries and have the bill being charged to their shared (with their partner) bank account. When paying for petrol and some sweets at the petrol station, the costs for the petrol are charged to the Petrol card account, while the cost of the sweets are charged to the debit amount stored on the card. For all of the purchases, the system automatically adds up the bonus points towards free airline tickets.*
>
> *While such a schema may have acceptance problems from a human point of view (technology threshold, privacy, security, etc.), it would be technically feasible when using smartcards. As a result, it may be the enabler of a new form of doing business and a new way of attracting customers.*
>
> *Finally, introducing such a one-card-does-it-all payment system would provide food for though for the existing banks and credit-card organisations. Thus far it has been their responsibility to issue credit-cards. With a one-card-does-it-all payment system this may change.*

This example shows how the development of a solution for a given problem may trigger the context in making further changes. It also illustrates the *enabling* effect that IT may have on an organisation. The second last paragraph of the example touches on the human issues. IT may be the enabler of the *one-card-does-it-all payment system* but lack of customer interest in, and acceptance of, such a payment system may be the disabler.

## 4.4 Planning horizon

The situation depicted in Figure 5 can be repeated on multiple levels of granularity depending on the planning horizon. This is illustrated in Figure 7. We have depicted three levels of architectural activities, corresponding to the strategic, tactical, and operational levels respectively. In reality, one may quite well recognise even more levels. At the top level (*Refocus*), we may be concerned with the development of a common (new) vision of the desired nature of an organisation's business, and the required IT. This project may give rise to more specific activities (*React*) that are aimed at realising some aspects of this strategy. At the bottom of the chain (*Realise*), we may find the projects that actually build the component-based solutions.

Each of the projects at a finer level of granularity are responsible for filling in part of the architecture as defined by the coarser grained higher-level architectures. At each of these levels of granularity, we will have to deal with the alignment between the business context and the defined solutions. The refocus, react, and realise activities will not be activities that are performed once only. A continuous cycling between refocus, react, and realise to meet new challenges and opportunities is probably a sign of a healthy situation.

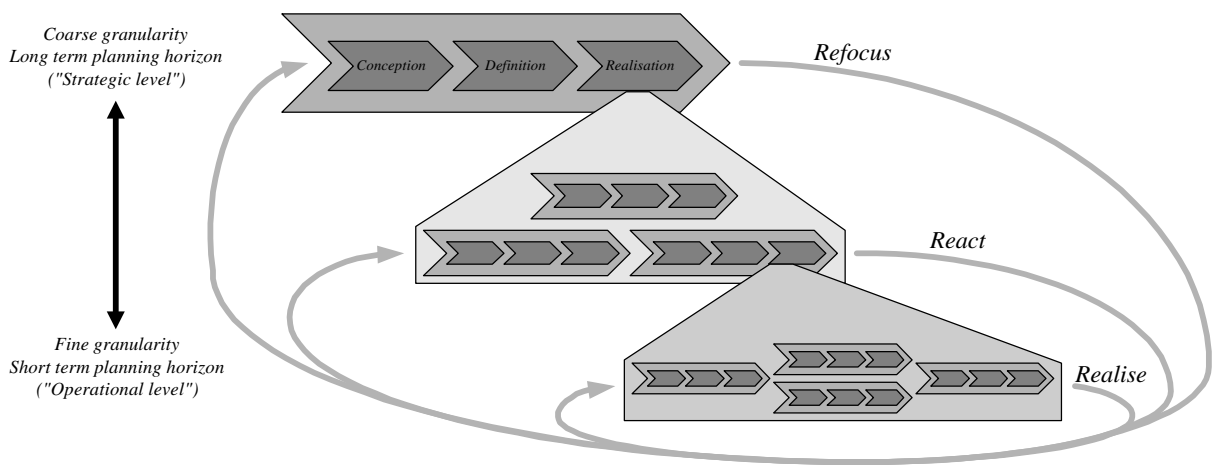

*Figure 7: Granularities and planning horizon*

## 5  Way of modelling

The aim of this section is to take a closer look at the way of modelling from an architecture-driven perspective. To this end we examine the architecture for the realisation of a business solution, the *business-solution architecture*, as a whole. We first try to provide an inventory on what areas need to be covered by such an architecture. We finish by briefly discussing the levels of granularity at which an architecture may be specified in relation to its intentions.

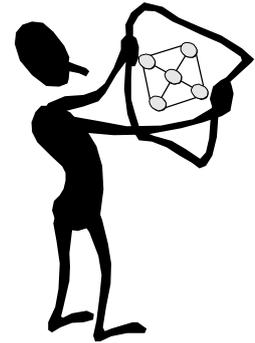

### 5.1  Solution definition

Consider the architecture of a new office building. In such an architecture we can focus on certain locations within the design. For example, the reception area, or the top floor. Alternatively, we could limit ourselves to one aspect of the architecture only. For instance the layout of the power-lines in a new building, or the highway infrastructure in a city plan. These are all examples of how to obtain what essentially are *sub-architectures*. Similarly, *sub-architectures* in an information processing architecture can be identified. By taking, a certain *view* on the architecture a sub-architecture follows.

Below we discuss two dimensions along which we can dissect business-solution architectures. The first dimension is concerned with obtaining a complete *definition* of the solution in its business context. The second dimension, is concerned with the actual realisation.

In [7], Tapscott identifies five essential views on business-solution architectures. Tapscott argues that the applications, the application view, are at the middle of a force field of four views. This is illustrated in Figure 8. These views essentially partition the elements of a business-solution according to the role they play in realising the business of the organisation. Below we discuss each of these views in more detail.

#### 5.1.1  Business view – What is our line-of-business?

The *business view* highlights *what* business is conducted by the organisation (or organisational sub-unit), in other words *'the line of business'*. This view considers an organisation as a service providing entity. The business view aims to describe only *what* an organisation does in terms of the services it provides. Details of *how* it does it are irrelevant at this stage. This is paralleled by Henderson and Venkatraman in [39] who make a strict distinction between an *external domain* and an *internal domain*:

> *The* external domain *is the business arena in which the firm competes and is concerned with decisions such as product-market offering and the distinctive strategy attributes that the differentiate the form from its competitors, as well as the range of "make-versus-buy" decisions, including partnerships and alliances.*
>
> *In contrast, the* internal domain *is concerned with choices pertaining to the logic of the administrative*

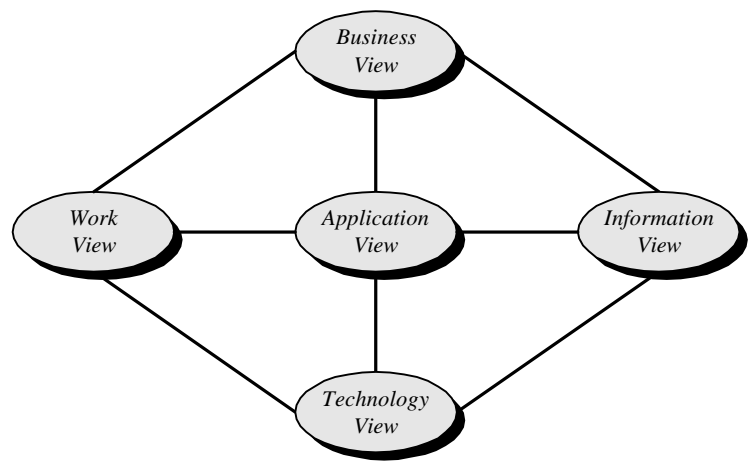

*Figure 8: Tapscott's views on business-solution architectures*

*structure (functional or divisional or matrix organisation) and the specific rationale for the design and redesign of critical business processes (product delivery, product development, customer service, total quality), as well as the acquisition and development of the human resource skills necessary for achieving the required organisational competencies.*

This view is therefore concerned with questions such as:

- Who are our suppliers?
- What is our product?
- Who are our customers?

Since we want to be *business driven*, and not *technology driven*, this part of the architecture is the *driver* for the other components of the architecture.

### 5.1.2 Technology view – What technological resources do we need to conduct our business?

The *technology view* focuses on the technology needed to *facilitate* the other components of the architecture. While the business view focuses on *what* an organisation does, this view focuses on *what with*. The technology view is the enabler of our business. The technology provides us with facilities for information processing that enable businesses [8]. Technological developments may, for example, lead to new distribution channels for services and goods. This will enable an organisation to conduct new forms of business. A typical example of this is the effect of 'electronic commerce' [7] and 'virtual communities' [42] as new distribution channels. Also the use of call-centres for direct sales is an example where new technology enables new forms of business activities.

### 5.1.3 Work view – How do we conduct our business?

The *work view* is concerned with the *how* of the business. This view is expressed in terms of work activities, associated resources, work locations, and needed information. In our context, an important goal for defining a work view is to determine the most effective ways in which the work activities can be supported by IT solutions. An important aspect of the work view is therefore the distinction between manual work and automated work.

Some essential questions that are addressed by the work view are:

- What are the processes by which we work?
- How do we organise ourselves?
- Which actors do the work?

Within the work view, it is also of particular importance to consider communication processes. Communication that is relevant for the business activities should be seen as a form of work. In the near future, applications will be used increasingly as a means for communication [12, 42]. Be it for internal use or communication with customers and suppliers. As an example, consider the work view of a hierarchical organisation as opposed to the one in a network organisation. One would expect the work processes in a hierarchical organisation to be highly detailed, while in a network organisation one would expect the focus to be on the facilitation of communication processes that enable the net*work*ing.

### 5.1.4 Information view – What information do we need to conduct our business?

The *information view* provides the information engineering perspective of business-solution architectures. The view is concerned with requirements for information resources. This will typically include a definition of what information will be stored, and what business-rule this should adhere to.

### 5.1.5 Application view – What information-processing do we use to conduct our business?

An *application view* describes the business realisation activities that will be automated. It defines how the automated parts of the work view will operate, which information resources are needed, and how technology will be used to achieve this. The application view is positioned in the centre of Figure 8 to emphasise the forces that bring about changes to this view. The dominant forces that will change the application view come from the business and the technology sides. Any changes to those views will directly or indirectly result in changes in the application view.

### 5.1.6 Solutions in context

The above five views also allow us to more precisely position the terms business-solution and IT-solution relative to each other. We have tried to do so in Figure 9. There the scopes of an IT-solution, and a business-solution are depicted in terms of the five views. What is stressed here is the fact that a business-solution should fit well within the context provided by the business.

A 'pure' IT-solution will generally only be part of the business-solution. Solving business issues typically requires more than hardware and software. New administrative processes, changes of the organisational structure, cultural changes, etc., are usually needed just as well. Software development methodologies in general are becoming more and more 'business aware'. For example, in [19], the Gartner Group already stated:

> *"By 1999, the market-leading software development methodologies will incorporate BPR methodologies (0.8 probability)."*

More recent, in [20], the Gartner Group states that:

> *"By 1999, the worlds of AD, workflow and business process re-engineering will merge, with AD*

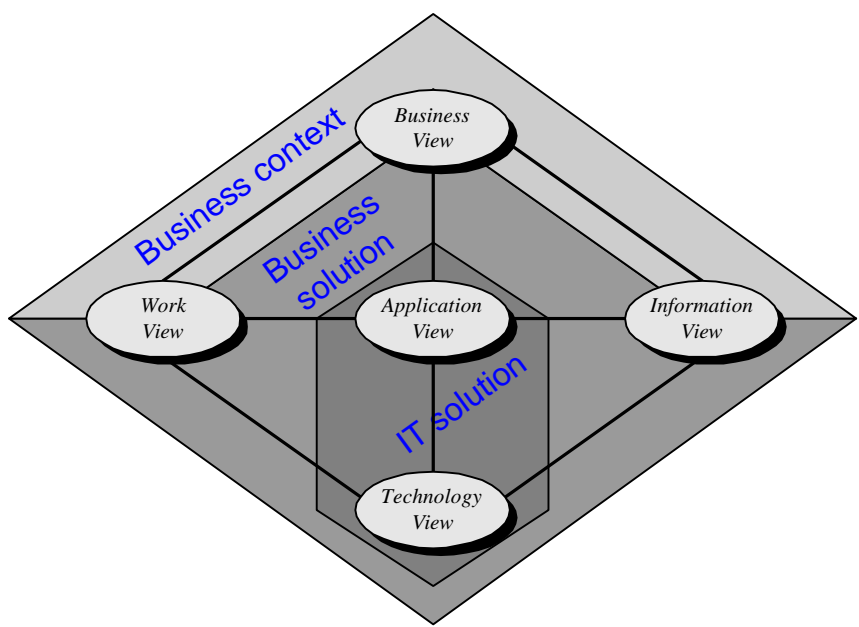

*Figure 9: Business solutions in context*

> *modelling tools, middle-ware, standards, objects and other emerging technologies serving as major rallying points (0.8 probability)."*

### 5.2 Solution realisation

The second dimension for splitting a business-solution architecture is based on the levels of detail regarding their physical *realisation*. In this dimension, two extremes can be distinguished: a *conceptual level* and a *physical level*. This applies to each of the five views in Figure 9.

At the *conceptual level* we are only concerned with architectural aspects that are completely independent of physical implementation details. This level is required to be completely independent of the underlying choices with respect to alternative implementation platforms and other implementation choices. This definition of conceptual level is in line with the one used in the ANSI/SPARC 3-level architecture as discussed in the ISO standard report on information system modelling [43].

As an example, consider dates. The fact that a date is represented as a string of 6(!?) characters is conceptually irrelevant. What *is* relevant is that (what we currently call) a date consists of a day, a month and a year.

Another example would be the use of a messaging mechanism between processes. The fact that when a bank customer does a credit-card application, the business process involved requests the client-credibility department to check the client's financial credibility, is a conceptual issue. What is not of conceptual relevance is whether this request is passed on via a database, an e-mail message, a memo, or a message on a message-queueing system.

It seems that the enabling effects of technologies for business, is usually best described at the conceptual level. Consider, for example, the Internet. The technology view of an architecture may express that the organisation's computer facilities are linked to the Internet and that WWW servers are available with access to databases. This is, however, purely at a physical level. What is more important for a business perspective is the enabling effect of this setup. The conceptual view of these pieces of physical technology is that it enables the business to automate the communication with clients and supplies, thus enabling new forms of business (electronic commerce). A business architect or business strategist will only be interested in this high level enabling effect of the underling technology.

At the *physical level* we are indeed concerned with the actual detailed work procedures, people, executables, and hardware. Between the conceptual and the physical level, there is a sort of a 'grey area'. This might be called the 'logical level', however, this is a sliding scale at which more and more physical details will be added.

When putting the two dimensions on a business-solution architecture together, the situation as depicted in

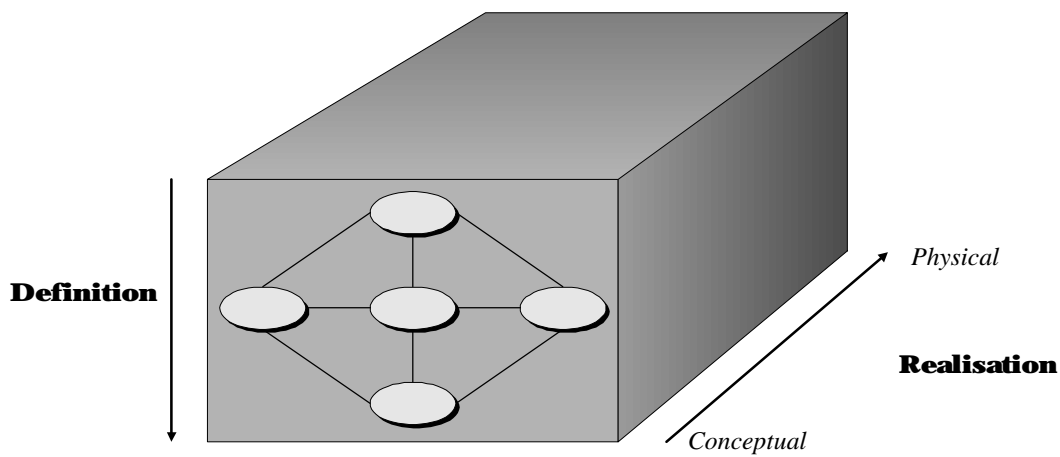

*Figure 10: Different dimensions of a business-realisation architecture*

Figure 10 results.

## 5.3 Granularity of architectures

In section 4.4, we already touched upon the issue that architectures were needed at different levels of granularity. It is very important to distinguish between the different levels of granularity needed. In [11], the Gartner Group argues that IT architectures fail more often than they succeed. The main reason for this seems to be the failure to distinguish between architectural 'blue-print' level issues and macrocosmic 'city planning' issues. See also Figure 11. This refers to the difference between a long-term architectural vision at a city planning level, and a detailed application architecture at the blue-print level. Even more, the long term architectural vision should not be too detailed. It is found to be practical to enforce a detailed blueprint for a whole set of applications since they are developed at different times by independent organisations. The guiding principle in this matter seems to be [11]:

> *Planning guidelines that affect the work of multiple development teams cannot be detailed and must focus on external interfaces and shared resources.*

There are at least two factors that influence the required levels of granularity. The first factor deals with the organisational positioning of the architecture (enterprise level vs. team level), and the second one with the planning horizon (short term vs. long term) of the architecture. Below we discuss these two factors in more detail.

The actual architecture itself is usually captured as a set of models, standards, and principles. These may all have a *descriptive* nature as well as *prescriptive* nature. For example, a city plan may *prescribe* that in a certain area only office buildings are to be built, that their height should not exceed 20 meters, and that ample parking space should be provided for. It may also *describe* the layout of the main infrastructure such as connecting roads, railroads, etc. For the IT domain some 'standard' principles can be found in [7, 8].

Depending on the planning horizon, an architectural document will be more or less detailed. A strategic architecture document should focus on principles and abstain from excessive use of models. In [44] the Gartner Group argues that, in particular when the audience is wider than IT professionals, a strategic architecture document should focus on principles and omit models. One important reason given is that models can quickly shift the focus towards tactics, instead of the original strategic objectives.

Architecture-driven development of business solutions can be defined as the process of arriving at an application by means of a sequence of architectures, where each of the successive architecture is more detailed. Each of the architectures in this sequence should of course be developed *under architecture*.

Architecture-driven application development acknowledges the fact that applications are developed in a context, a context that will usually be provided as a prescriptive (long term) architecture (analogous to a zoning plan). This is what sets *architecture-driven* development apart from *model-driven* development. Models will still play a crucial role in architecture-driven application development, but what is just as important is the context in which the development occurs and the applications needs to fit. Architecture-driven development stresses the need for 'head up' development rather than 'head down' development; i.e. under architecture.

### 5.3.1 Organisational context

The organisational positioning of the architecture acknowledges the fact that there will usually be a difference in granularity between e.g. an enterprise-wide architecture and the one that focuses on the activities of a single team. An enterprise-wide architecture will usually be coarse grained and focus on general infrastructural issues. One would also expect such an architecture to be more of a *pre*scribing nature rather than a *de*scribing nature; i.e. principles rather than models. An architecture concerning a single team tends to be more fine grained and deal with all relevant aspects. In [11], a distinction is made between the '*city planning*' level and the '*blue-print*' level.

As an example, suppose the board of directors of an organisation decides to use a certain hardware platform as the standard for the IT activities in their organisation. Then this decision should be part of the technology view at the 'city planning' level. On the other hand, a detailed definition of how a bank clerk should evaluate the credibility of a client is part of the work view at the 'blue-print' level.

The overall effect is that, specific to the hierarchy of the organisation, a hierarchy of architectures follows. We have illustrated this in Figure 12. An architecture for the departmental level should be *in line with* the more coarse grained architecture at the enterprise level. At the same time, the team level architectures should all be in line with their relevant departmental architectures. Note that for this discussion it is irrelevant how a particular organisation has been structured, be it by means of departments, sub-departments, teams, or logical service units.

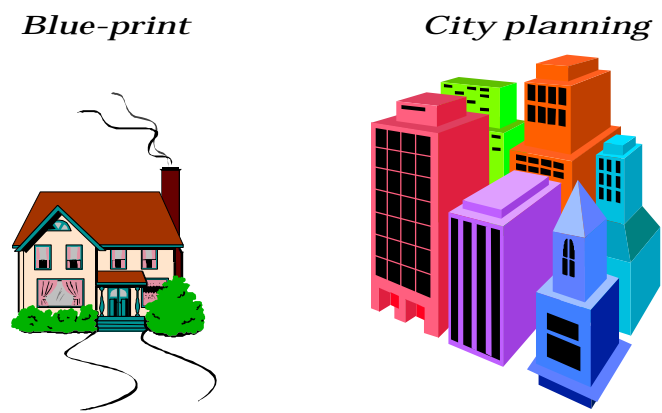

*Figure 11: Different levels of architecture*

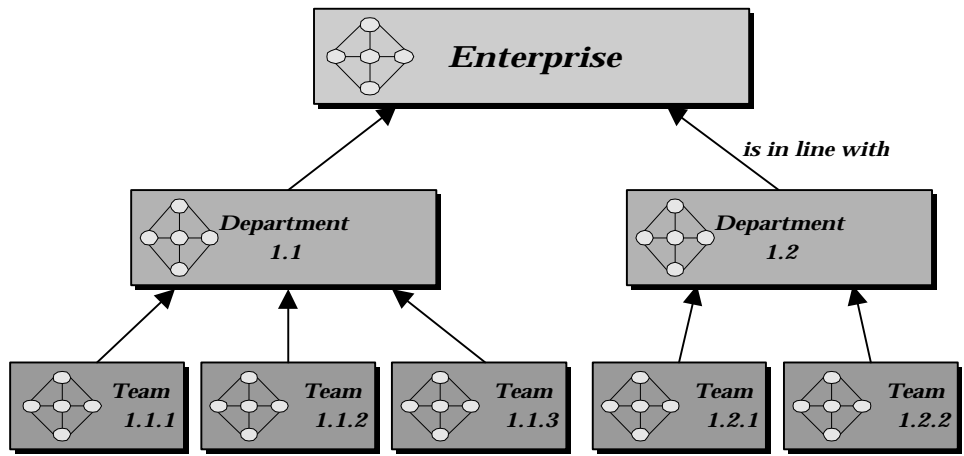

*Figure 12: Granularity levels for business-realisation architectures*

When developing a business-solution architecture from scratch, one will have the luxury of being able to define the architecture top-down. In practice, however, the lower levels of granularity may already exist. For example, existing hardware and legacy applications. The existence of these 'legacy' pieces of architecture will usually prohibit a clean top-down approach. For instance consider what happens if two companies, and their respective IT architectures, have to be merged.

### 5.3.2 Planning horizon

The planning horizon is the second factor that influences the granularity of an architecture. In section 4.4, we already discussed some of the issues involved. In Figure 7, we showed three levels of architectural activities: *refocus* on the strategic level, *react* on the tactical level, and *realise* on the operational level. We also mentioned that these activities should be cyclic in nature to constantly adapt to changes in the business environment.

An architecture on the strategic level should provide a long term *vision* of the direction in which the actual (operational) architecture should evolve. This architecture should serve as a *reference point* for new developments, and will usually focus on prescribing elements. When moving via the tactical level to the operational level, describing elements will become more predominant.

## 6 Way of working

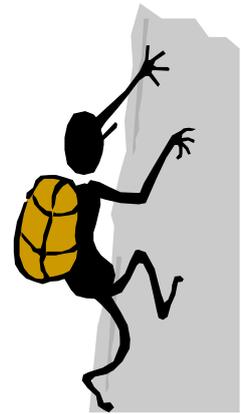

In this section, we discuss some of the issues involved the way of working when developing architecture-driven business solutions. We revisit some of DAVINCI's key drivers and discuss ways to operationalise these drivers in practice.

### 6.1 Architectural aims

Architecture projects may be initiated for different reasons. There seem to be three orthogonal reasons for doing this, which together define a kind of 'vector space'. A given project may focus solely on one of these aims, or it may aim at a combination of them. Depending on the project aim, a different variation of the way of working may be required. The three base aims are:

1. *Realign.*

   **What:** Alignment of business and IT.

   **Aim:** Improvement of current business-realisation activities, and better utilisation of IT resources in line with the vision architecture.

   **Why:** Bad alignment of business and/or between business and IT.

2. *Refit.*

   **What:** Replacement of the underlying technology.

   **Aim:** Improvements of or changes to the underlying technologies used, in line with the vision architecture.

   **Why:** Preparing for the future; decommissioning ill-performing technology.

3. *Renew.*

   **What:** Development of new applications due to new business requirements.

   **Aim:** Improvements of or changes to the current architecture to cater for new business requirements, in line with the vision architecture.

   **Why:** New business or changes to the existing business require new business processes and support.

In parallel to all of these possible activities, a renaissance process is needed. Before starting a development process under architecture, awareness needs to be created among the people involved. They need to be made aware, and convinced, of the benefits of architecture-driven development This means that the involved people need to go beyond their current beliefs and dogmas surrounding system development. A paradigm shift [7] is needed. The renaissance required, obviously involves some serious change management [45]. The renaissance process also requires attention during the intake phase of projects. If during an intake it becomes clear that the organisation will not accept the new ideas, then the conclusion may have to be that they are simply not ready for architecture-driven application development.

### 6.2 Driven by needs; enabled by technology

In trying to operationalise the 'Driven by needs; enabled by technology' driver, it is important to establish a way to accommodate the alignment processes as depicted in sections 4.1 and 4.3. An important role in this is played by the language used to communicate ideas and reach consensus. A language is needed that allows business and IT people to communicate

Tapscott [7] proposed *architecture principles* as a way to provide these handles of control. In his terms, architecture principles are statements of *preferred architectural direction or practice*. Keen [8] refers to these principles as *policies*.

Architecture principles are in essence constraints on the IT (or business) architecture of an organisation. For example, an architecture principle could be defined as:

> *Our systems should utilise standard, shareable, reusable components across the enterprise.*

This statement constraints the IT architecture of this particular organisation since the use of a non-sharable component, say, in the IT architecture would constitute a violation of this architectural constraint. A collection of architecture principles can be viewed as collectively defining the preferred (or expected) direction in which the architecture should evolve. As such, the principles provide a context for any further architectural design decisions.

Architecture principles can be employed as a common language between business and IT people. Doing so allows architecture principles to play the following key roles in a development process:

- They provide handles of control. The resultant is that architecture decisions become more controllable by both business and IT management.
- Design decisions can be motivated by referring to architecture principles and making trade-offs between principles. This enables traceability and impact analysis.
- They eliminate the need for evaluating endless alternatives in the modelling and programming stages by agreeing up front on preferred directions.

In [7, 8, 46] criteria for good architecture principles can be found. Based on these sources, the following list of criteria can be compiled:

1. *Each principle clearly states a fundamental belief of the organisation.*

    The organisation should be committed to them.

2. *Principles should express a (preferred) direction.*

    For example, "information is an asset" doesn't express a direction, while "our IT architecture should be

3. *They should be simply stated and understandable to both business and IT managers.*

4. *Principles should be rationalised. Preferably in terms of 'higher order' architecture principles.*

    For example: "Our IT architecture should be flexible *because* we want to be product innovators". In addition, more motivation may be provided such as: Why did this principle get stated this way? What alternatives were discussed?

5. *The implications need to be discussed and documented.*

    For example, what impact does this principle have on the IT organisation? On management processes? On technology?

An example architecture principle, taken from [46], architecture principle can be found in Figure 13. In the rationale of this example principle we can find the following higher order principle:

> "… and IT must be better able to build flexibility into its systems and allow them to adapt to changing business requirements."

This can be more compactly re-formulated as:

> *Our IT architecture should be flexible enough to adapt to changing business requirements.*

The rationalisation of architecture principles in terms of higher order architecture principles allows us to compose principle-chains. An example of such a principle-chain is given below.

1. *Operational-excellence has our highest priority.*

    This is a basic axiom of our organisation's strategy.

2. *Our production process should have a high level of continuity.*

    One of the ways in which we have chosen to realise operational-excellence (besides cost savings, and faster production times without quality reductions), is to ensure the continuity of the actual production process.

3. *We need to make use of proven technology for our IT architecture.*

    As large parts of our production process depend on IT components, these components should be highly reliable. The use of proven technology is one way to warrant this.

> **Principle**
>
> *Our systems should utilise standard, shareable, reusable components across the enterprise.*
>
> **Rationale**
>
> It is critical that the IT organisation improve its response time to business needs and delivery systems faster and with better quality. Our organisation is going through substantial change and IT must be better able to build flexibility into its systems and allow them to adapt to changing business requirements.
>
> Using standard components as the basis for defining and building the architecture and delivered systems can improve our productivity by using previously defined and built components. Rather than build new components each time, developers can concentrate on new business requirements, rather than redoing existing work. We believe that the ability of our systems to adapt to changing requirements can be improved by using standard components.
>
> **Implications**
>
> There are a number of management and organisational implications from this principle:
>
> - A means of co-ordinating, defining, and communicating the available standard components will need to be developed.
> - Areas where definitions of standard components will be required include business processes, applications (at all levels), and technology components (processors, system software, network components, languages and development tools, and data, such as subject databases, conceptual designs, physical implementations, etc.).
>
> A management process will be required to track the generation and usage of these shareable components and to standardise them where needed.
>
> - A standard definition of each component type will also need to be defined. This could be facilitated through a well-implemented common system delivery methodology.
> - A library of definitions, terms, access rules, characteristics, and interrelationships of each of the application, information, technology and, potentially, organisational and business components needs to be implemented corporate wide.

*Figure 13: Example architecture principle*

This example also illustrates a possible trade-off between principles. The use of proven technology may result in less flexibility of the IT architecture, which may contradict principles such as:

- *Our IT architecture should be flexible enough to adapt to changing business requirements.*
- *We should be able to quickly adapt to the production of new products.*

In [46] a sketch of a procedure to obtain architecture principles can be found.

## 6.3 Results oriented; not role oriented

In trying to steer away from the potential problems involved in a role oriented software development process, we propose the notion of *persistence of accountability.* The idea is to make all people involved accountable for the end-result of a development process. There needs to be a feeling of being 'accountable'. Consider, for instance, a traditional analysis, design, and implementation phase. It is the responsibility of designers to ensure the results of the analysis phase is sufficient for them to start their work. Conversely, it is the analysts' responsibility to ensure that the design fits the bill. In short:

> To the designer:
>
> *You designed it. You make sure it is built that way!*
>
> To the constructor:
>
> *You'll have to build it. You make sure the design is right!*

A similar shared accountability applies to designers and implementers. The need to look beyond the traditional boundaries of the distinct development phases and associated teams was already identified in publications such as [47, 48]. One may even argue that system management and maintenance should be included as well. In other words, the designers and implementers should be held accountable for the maintainability of the system.

In the context of ABS, persistence of accountability will also be used to bridge the divide between business and IT, in particular business consultants, business architects, and IT architects. IT should be an enabler, not a driver. To be an enabler, IT architects should be a sparring partner of business architects and consultants. Accountability should be shared. This view is essentially also supported by [39], who argue that a business strategy and an IT strategy should be developed in tandem.

## 6.4 Evolution is a constant

Most applications developed using an architecture-driven approach will need to be future-proof. This does not mean that applications need never be changed. What it does mean is that changing them does not entail a redesign or re-implementation of the entire application. This guarantee is met by ensuring that the business-solution architecture, or at least the automated parts thereof, is inherently flexible. This means that making changes becomes easier; cheaper. Some key principles in this context are:

- *Conceptualisation principle.* By first focusing on the *what* issues, we obtain specifications and insights that are independent of the underlying choices with respect to implementation in terms of alternative forms of physical technology. Doing this makes the architectures, and implementations, less susceptible to the effects of technological changes. This principle is borrowed from [43].

- *Change absorption.* The cheapest and quickest way to deal with changes is by not making changes. Therefore, where possible, architectures should be made highly configurable. For applications the effect will be that changes within a certain flexibility-bandwidth are easy to make (ideally by end-users, but at least by power-users) without changing the actual code. In other words, changes are absorbed by the application as is. This allows the 'shocks' of change to be absorbed by the application 'as is' without the need to change the actual code. ERP software packages are typical examples of highly configurable IT solutions.

- *Change encapsulation.* By grouping elements together that form a cohesive group, and minimising dependencies between different groups (e.g. by information hiding), effects of changes can be limited. In other words, changes are encapsulated. If changes do need to be made, for instance if they cannot be absorbed, the effects of these changes should remain as local as possible; i.e. they are encapsulated.

Yielding a flexible architecture requires an understanding of what the causes for changes of an existing architecture are. As our main concern is application development, we first focus on the automated parts of the architecture. This will usually be implemented in terms of some applications (e.g. comprising an *information*

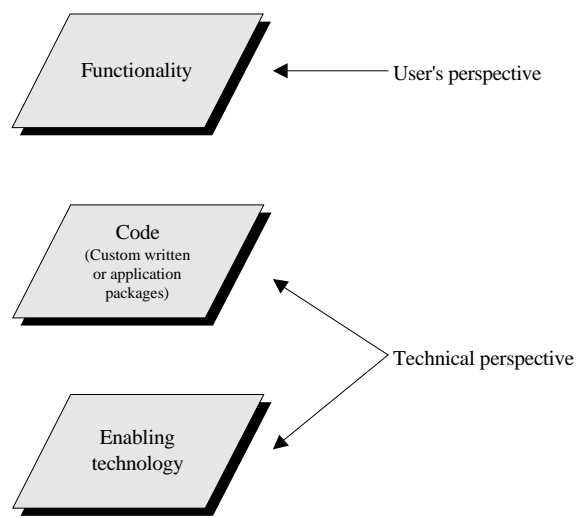

*Figure 14: The three essential parts of an information system*

*system*). These applications will have to deal with a number of changes in the course of their life cycle,. These changes can come from two main directions: the functionality side, and the technology side. To illustrate this, consider Figure 14, which is taken from [9]. As argued in [9], there are three essential parts to an application. From a user's perspective, an information system needs to provide certain *functionality*. This functionality is ultimately provided by the *code* of the actual system, which consists of a combination of custom-written or acquired applications. This code sits on top of the lower levels of software and hardware, the *enabling technology*.

The *code* is obviously the most expensive part of an application. As can be seen in this figure, changes to this code come from two directions. There is a potential *pull* for change in the form of changing functional requirements and at the same time a potential *push* for changes by developments in the underlying enabling technology. The pace of changes in functionality will increase as the organisation moves through the levels discussed in Figure 1.

Applying old and tested software engineering practices [13, 28] can reduce the impact of changes to the code. By modularising the code and by introducing abstraction layers, we can obtain change encapsulation. To further illustrate this point, the code of an application can roughly be grouped into five main layers:

1. *Interaction Layer*. This layer is concerned with interactions. For example, user dialogues.
2. *Service Layer.* This layer provides the services offered by the organisation to the outside world. One may also choose to call this the 'workflow' layer. One may even go as far and claim that this is laer corresponds to the automated part of the business view.
3. *Process Layer*. This is the 'home' of the automated business processes. This is where the real '
    done. Therefore, one may say that this corresponds to the automated parts of the work view.
4. *Information Layer*. This is where we can find the information needed to do the work. Typical functionality in this layer will be concerned with basic actions to create, update, or delete objects, and the enforcement of business rules.
5. *Resource Layer*. This is where we will find databases, interfaces to legacy applications, etc.

The distinction between a service, process, and information layer is inspired by the work in [Barros, 1997 #121; Barros, 1997 #1567]. The complete set of layers is illustrated in Figure 15. This figure is based on the technical reference architecture as presented in [4, 5, 49], where we have introduced some extra layers. In this architecture, the service layer, process layer, and information layer are comprised of so called *business-objects*. A business-object is a component that provides some business functionality. They are typically more fine grained than shrink-wrapped application packages. Business-objects are found to have a positive effect on the flexibility of applications [51].

The use of these five layers of abstraction is only a brief illustration of what can be achieved by a component (or business-objects) based approach to system design and implementation. Each layer can be divided in further sub-layers of specificity and synergy. Over the next 4 years, component-based application development is expected to become increasingly a mainstream activity. This is supported by strategic statements from the Gartner Group [20]:

> *"Object and component environments will evolve to support flexible packaging and increased reuse, starting with graphical user interface and technical hierarchies and components through 2001 (0.7 probability)."*

> *"Through 2001, packaged-software vendors will face increasing customer pressure to deliver functionality in the form of objects or templates using widely available AD environments (0.8 probability)."*

> *"By 2001, 60 percent to 70 percent of all new applications will either be assemblies of business-objects, customisations of templates or both, increasing the ability to cope with changes (0.7 probability)."*

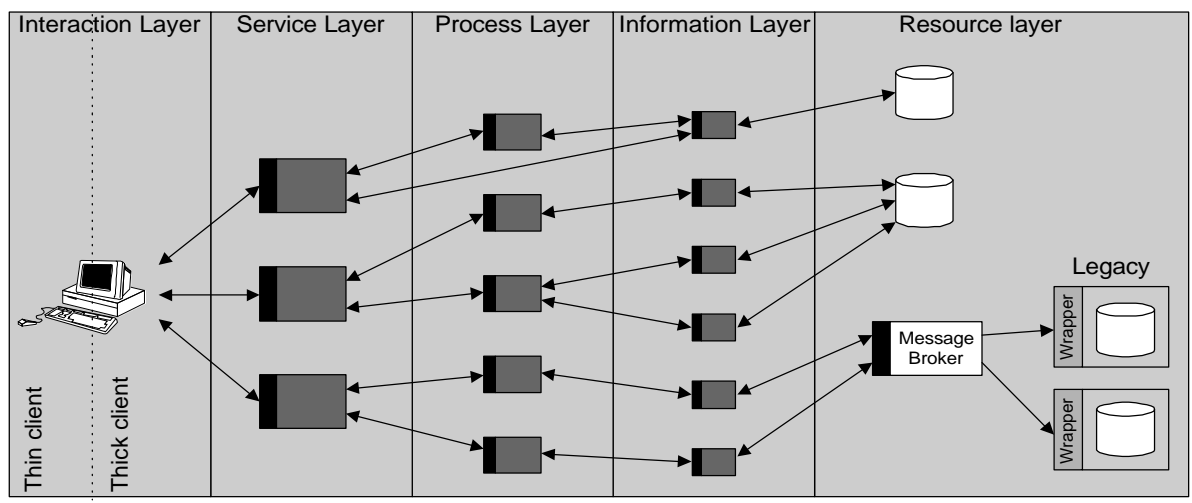

*Figure 15: Encapsulating change by abstraction layers*

In [50] they furthermore state:

> *"By 2001, more than three-quarters of new applications will be built, in part, using pre-existing components, without direct exposure to low-level procedural middle-ware APIs (0.7 probability)."*

The above discussion on components and business objects clearly focuses on the automated parts of the business (information) processing in an organisation. However, similar considerations will apply to manual activities. Modularization of manual work by making a distinction between technology specific, general business, and business-service specific activities may enable flexibility of manual work as well.

## 7 Conclusion

In this report we have presented architecture-driven business solutions (ABS) as a way of thinking for the development of business solutions. We have argued the need for ABS by listing and motivating its key drivers. Besides defining the ABS way of thinking, we have discussed some of the consequences for the way of working and the way of modelling of methods and techniques used in an ABS context.

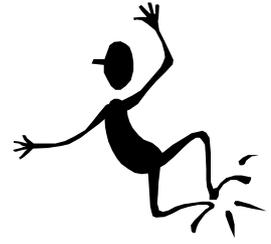

For Origin it is now time to develop appropriate methods and techniques to fill the ABS toolbox. DAVINCI only provides a theoretical backdrop. To make things work, more concrete working procedures, techniques, tooling, etc, are essential. People need to be selected and trained for the different activities involved in ABS. This is only the beginning!

8    Definitions

**ABS** – Architecture-driven business solutions.

**AD²** – See architecture-driven application development.

**Architecture** – Style + discipline.

**Architecture-driven application development** – The process of arriving at an application by means of a sequence of architectures, where each successive architecture is more detailed. Each of the architectures in the sequence needs to be is *developed under architecture*.

**Architecture-driven business solutions** – The process of arriving at business solutions, by means of a sequence of architectures, where each successive architecture is more detailed. Each of the architectures in the sequence needs to be *developed under architecture*. The solutions may involve both automated and manual aspects.

**Architecture principles** – Statements of preferred architectural direction or practice.

**Application view –** That part of a business-solution architecture, which tries to answer the question: *What information-processing do we use to conduct our business?*

**BAS** – Business Application Solutions, an Origin service line.

**Business-solution architecture –** The architecture of a business solution, which fits in the context of a given context architecture.

**Business view** – That part of a business-solution architecture, which tries to answer the question: *What is our line-of-business?*

**Context architecture** – When developing a new architecture, this architecture will have to fit within a certain architectural context; the *context architecture*. This context may be the current context of the applications or business under development, but it can also be a long term vision. Compare this to the way the architectural design of a new office building fits within the zoning plan of the area in which the building will be erected.

**DA VINCI** – Origin's way of thinking with regards to ABS.

**Development under architecture –** The development of solutions that fit within the context as defined by some context architecture. The context architecture gives guidance to the development process.

**Information view** – That part of a business-solution architecture, which tries to answer the question: *What information do we need to conduct our business?*

**Strategic alignment –** The alignment between business and IT strategies.

**TRA** – See technical reference architecture.

**Technical reference architecture** – A reference architecture for component based transactional solutions, focussing on:
1. a topology of layers into which the internal components of a logical component are divided, and the responsibilities of the components within these layers.
2. a detailed definition of the way in which logical components interact with legacy applications using message brokers.

**Technology view** – That part of a business-solution architecture, which tries to answer the question: *What technological resources do we need to conduct our business?*

**Way of controlling** – The managerial aspects of a development method.

**Way of enacting –** A combination of a way of working and way of modelling.

**Way of modelling** – A description of the concepts in terms of which we define (prescribe or describe) the architectures.

**Way of learning –** A description of how the development organisation may learn (and the method refined) based on experiences in using the method in practice.

**Way of planning** – The way in which the enactment and learning phases of the development is planned.

**Way of supporting** – The support that may be provided by (possibly automated) tools.

**Way of thinking** – The way of thinking verbalises the assumptions and viewpoints of the method on the kinds of problem domains, solutions and modellers.

**Way of working** – The actual process of developing a system. It defines the possible tasks, including sub-tasks, and ordering of tasks, to be performed as part of the development process. It also provides guidelines and suggestions (heuristics) on how these tasks should be performed.

**Work view** – That part of a business-solution architecture, which tries to answer the question: *How do we conduct our business?*